\documentclass[structabstract,longauth]{aa}

\usepackage{url}
\usepackage{twoopt}
\usepackage[english]{babel}          

\usepackage{natbib}
\bibpunct{(}{)}{;}{a}{}{,} 

\usepackage[utf8]{inputenc}
\usepackage[normalem]{ulem}
\usepackage{siunitx}
\usepackage{amsmath, amssymb}
\usepackage[farskip=0pt]{subfig}

\usepackage{multirow,bigstrut,ctable}
\usepackage[normalem]{ulem}
\usepackage{rotating}
\usepackage[varg]{txfonts} 
\usepackage{threeparttable}

\def \deg         {\text{$^{\circ}$}}
\def \arcmin      {\text{$^\prime$}}
\def \arcsec      {\text{$^{\prime\prime}$}}
\def \hour        {$^{\mathrm{h}}$}
\def \min         {$^{\mathrm{m}}$}
\def \sec         {$^{\mathrm{s}}$}
\def \mjybeam     {mJy\,beam$^{-1}$}

\newcommand{\hms}[3]{{#1}\hour{#2}\min{#3}\sec}
\newcommand{\dms}[3]{{#1}\deg{#2}\arcmin{#3}\arcsec}
\newcommand{\beam}[2]{{#1}\arcsec$\times${#2}\arcsec}

\begin{document}

\title{Cassiopeia A, Cygnus A, Taurus A, and Virgo A \\at ultra-low radio frequencies
}
\titlerunning{A-team at ultra-low radio frequencies}

\author{
F. de Gasperin \inst{ 1 }
J. Vink \inst{ 2,3,4 }
J.P. McKean \inst{ 5,6 }
A. Asgekar \inst{ 6,27 }
M.J. Bentum \inst{ 6,7 }
R. Blaauw \inst{ 6 }
A. Bonafede \inst{ 14,15,1 }
M. Br\"uggen \inst{ 1 }
F. Breitling \inst{ 24 }
W.N. Brouw \inst{ 5,6 }
H.R. Butcher \inst{ 26 }
B. Ciardi \inst{ 28 }
V. Cuciti \inst{ 1 }
M. de Vos \inst{ 6 }
S. Duscha \inst{ 6 }
J. Eisl\"offel \inst{ 11 }
D. Engels \inst{ 1 }
R.A. Fallows \inst{ 6 }
T.M.O. Franzen \inst{ 6 }
M.A. Garrett \inst{ 18,8 }
A.W. Gunst \inst{ 6 }
J. H\"orandel \inst{ 32,33,34 }
G. Heald \inst{ 25 }
L.V.E. Koopmans \inst{ 5 }
A. Krankowski \inst{ 23 }
P. Maat \inst{ 6 }
G. Mann \inst{ 24 }
M. Mevius \inst{ 6 }
G. Miley \inst{ 8 }
A. Nelles \inst{ 21,22 }
M.J. Norden \inst{ 6 }
A.R. Offringa \inst{ 6,5 }
E. Orr\'u \inst{ 6 }
H. Paas \inst{ 46 }
M. Pandey-Pommier \inst{ 30,31 }
R. Pizzo \inst{ 6 }
W. Reich \inst{ 35 }
A. Rowlinson \inst{ 2,6 }
D.J. Schwarz \inst{ 29 }
A. Shulevski \inst{ 2 }
O. Smirnov \inst{ 9,10 }
M. Soida \inst{ 12 }
M. Tagger \inst{ 13 }
M.C. Toribio \inst{ 17 }
A. van Ardenne \inst{ 6 }
A.J. van der Horst \inst{ 19,20 }
M.P. van Haarlem \inst{ 6 }
R. J. van Weeren \inst{ 8 }
C. Vocks \inst{ 24 }
O. Wucknitz \inst{ 35 }
P. Zarka \inst{ 16 }
P. Zucca \inst{ 6 }
}
\authorrunning{F.~de~Gasperin et al.}

\institute{ 
Hamburger Sternwarte, Universit\""at Hamburg, Gojenbergsweg 112, 21029, Hamburg, Germany \email{fdg@hs.uni-hamburg.de}
\and Anton Pannekoek Institute for Astronomy, University of Amsterdam, Science Park 904, 1098 XH Amsterdam, The Netherlands
\and GRAPPA, University of Amsterdam, Science Park 904, 1098 XH Amsterdam, The Netherlands
\and SRON, Netherlands Institute for Space Research, Utrecht, The Netherlands
\and Kapteyn Astronomical Institute, University of Groningen, PO Box 800, NL-9700 AV Groningen, the Netherlands
\and ASTRON - the Netherlands Institute for Radio Astronomy, Oude Hoogeveensedijk 4, 7991 PD Dwingeloo, the Netherlands
\and Eindhoven University of Technology, P.O. Box 513, NL-5600 MB Eindhoven, The Netherlands
\and Leiden Observatory, Leiden University, Niels Bohrweg 2, NL-2333CA, Leiden, The Netherlands
\and Department of Physics \& Electronics, Rhodes University, 6139, Grahamstown, South Africa
\and South African Radio Astronomy Observatory, 7925, Observatory, Cape Town, South Africa
\and Thüringer Landessternwarte, Sternwarte 5, D-07778 Tautenburg, Germany
\and Astronomical Observatory, Jagiellonian University, ul. Orla 171, 30-244 Krak\'ow, Poland
\and LPC2E - Universit\'e d’Orl\'eans/CNRS, Orl\'eans, France
\and Dipartimento di Fisica e Astronomia, Universit\'a di Bologna, via P. Gobetti 93/2, 40129, Bologna, Italy
\and INAF - Istituto di Radioastronomia, Bologna Via Gobetti 101, I–40129 Bologna, Italy
\and LESIA \& USN, Observatoire de Paris, CNRS, PSL, SU/UP/UO, 92195 Meudon, France
\and Department of Space, Earth and Environment, Chalmers University of Technology, Onsala Space Observatory, SE-439 92 Onsala, Sweden
\and Jodrell Bank Centre for Astrophysics, Department of Physics \& Astronomy, The University of Manchester, Alan Turing Building, Oxford Road, Manchester, M13 9PL, UK
\and Department of Physics, The George Washington University, 725 21st Street NW, Washington, DC 20052, USA
\and Astronomy, Physics, and Statistics Institute of Sciences (APSIS), The George Washington University, Washington, DC 20052, USA
\and Erlangen Center for Astroparticle Physics (ECAP), Friedrich-Alexander-Universität Erlangen-Nürnberg, 91058 Erlangen, Germany
\and DESY, Platanenallee 6, 15738 Zeuthen, Germany
\and Space Radio-Diagnostics Research Centre, University of Warmia and Mazury in Olsztyn, Prawochenskiego 9, 10-720 Olsztyn, Poland
\and Leibniz-Institut für Astrophysik Potsdam (AIP), An der Sternwarte 16, 14482 Potsdam, Germany
\and CSIRO Astronomy and Space Science, PO Box 1130, Bentley WA 6102, Australia
\and Research School of Astronomy and Astrophysics, Mount Stromlo Observatory, Cotter Road, Weston Creek, ACT 2611 Australia
\and Shell Technology Center, Bangalore 562149, India
\and Max Planck Institute for Astrophysics, Karl-Schwarzschild-Str. 1, 85748, Garching, Germany
\and Fakult\""at f\""ur Physik, Universit\""at Bielefeld, Postfach 100131, 33501 Bielefeld, Germany
\and USN, Station de Radioastronomie de Nançay Observatoire de Paris route de Souesmes 18330 Nançay France
\and Univ Lyon, Univ Lyon1, Ens de Lyon, CNRS, Centre de Recherche Astrophysique de Lyon UMR5574, 9 av Charles André F- 69230, Saint-Genis-Laval France
\and Department of Astrophysics/IMAPP, Radboud Universiteit, P.O. Box 9010, 6500 GL Nijmegen, The Netehrlands
\and Nikhef, Science Park 105, 1098 XG Amsterdam, The Netherlands
\and Vrije Universiteit Brussel, Physics Department, Pleinlaan 2, 1050 Brussels, Belgium
\and Max-Planck-Institut f\""ur Radioastronomie, Auf dem H\""ugel 69, 53121 Bonn, Germany
}

\date{Received ... / Accepted ...}

\abstract
{The four persistent radio sources in the northern sky with the highest flux density at metre wavelengths are Cassiopeia A, Cygnus A, Taurus A, and Virgo A; collectively they are called the A-team. Their flux densities at ultra-low frequencies ($<100$~MHz) can reach several thousands of janskys, and they often contaminate observations of the low-frequency sky by interfering with image processing. Furthermore, these sources are foreground objects for all-sky observations hampering the study of faint signals, such as the cosmological 21 cm line from the epoch of reionisation.}
{We aim to produce robust models for the surface brightness emission as a function of frequency for the A-team sources at ultra-low frequencies. These models are needed for the calibration and imaging of wide-area surveys of the sky with low-frequency interferometers. This requires obtaining images at an angular resolution better than 15\arcsec\  with a high dynamic range and good image fidelity.}
{We observed the A-team with the Low Frequency Array (LOFAR) at frequencies between 30 MHz and 77~MHz using the Low Band Antenna (LBA) system. We reduced the datasets and obtained an image for each A-team source.}
{The paper presents the best models to date for the sources Cassiopeia A, Cygnus A, Taurus A, and Virgo A between 30 MHz and 77~MHz. We were able to obtain the aimed resolution and dynamic range in all cases. Owing to its compactness and complexity, observations with the long baselines of the International LOFAR Telescope will be required to improve the source model for Cygnus A further.}
{}


\maketitle

\section{Introduction}
\label{sec:introduction}

Historically, the radio sources with the highest flux density in the sky were named after the constellation in which they were found followed by a letter starting with ``A''. They were then grouped in the so-called A-team\footnote{This is also a famous TV series from the 1980s.}. In this work, we focus on the four persistent radio sources with the highest flux density (below GHz frequency) in the northern sky: Cassiopeia~A, Cygnus~A, Taurus~A, and Virgo~A (see Table~\ref{tab:targets}), which are all very different in nature. Cassiopeia A is a prototypical supernova remnant, while a large fraction of the radio emission from Taurus A is powered by the central Crab pulsar and its associated shocked pulsar wind; Cygnus A is a very powerful Fanaroff-Riley (FR) type-II radio galaxy at the centre of a massive, merging galaxy cluster \citep{Markevitch1999}; and Virgo A is an amorphous radio source powered by a black hole with mass $M_{\rm BH} = (6.5 \pm 0.7) \times 10^9$\,M$_\odot$ \citep{Akiyama2019} at the centre of a small, nearby galaxy cluster. Cygnus A is at the distance of 232 Mpc ($z=0.056$) and its radio power is $L_{1.4\ \rm GHz} \simeq 1.2 \times 10^{28}$ W~Hz$^{-1}$ (assuming a flat $\Lambda$CDM cosmology with $H_0=71$~km\,s$^{-1}$~Mpc$^{-1}$ and $\Omega_{\rm M}=0.27$), which is among the highest registered for radio galaxies. Virgo A is at the centre of the closest galaxy cluster at the distance of 16.5 Mpc ($z=0.00428$) and its radio luminosity is $L_{1.4\ \rm GHz} \simeq 8.3 \times 10^{24}$ W~Hz$^{-1}$. Cassiopeia A and Taurus A are Galactic sources at the distance of 3.4~kpc \citep{Reed1995} and $\sim3$ kpc \citep{Bailer-Jones2018}, respectively.

These bright objects present a challenge for the calibration of radio interferometers, as their emission can leak into the primary beam side lobes and corrupt the dataset \citep[e.g.][]{Patil2017a}. This is especially relevant for low-frequency phased arrays, where the side lobes are less suppressed compared to dish-based instruments. A number of analysis techniques have been developed to account for the effect of the A-team in the data. A possibility is to predict the time--frequency regions of the observation where one side lobe of the beam crosses one of the A-team sources \citep{Shimwell2017}. If the predicted contaminating flux density is above a certain threshold, then that part of the data is discarded. This procedure is usually fast and it has been proven to be robust for observations with the High Band Antenna (HBA) system of the Low Frequency Array \citep[LOFAR;][]{VanHaarlem2013}, but it requires an accurate modelling of the primary beam side lobes. In the case of HBA observations, the amount of data loss is typically 5 to 10 \%. Another technique that has been developed is the so-called demix \citep{VanderTol2009}. This technique requires high-frequency and time resolution data and it is conceptually similar to the ``peeling'' process \citep{Noordam2004}. The dataset is phase-shifted towards the direction of the A-team source and is averaged down in time and frequency to smear all other sources. A calibration is then performed against a pre-existing model. Then, the model visibilities of the A-team source, corrupted with the solutions just obtained, is subtracted from the full-resolution dataset. When the A-team source is very close to a given target field ($<30\deg$), a standard peeling \citep{Noordam2004} or a multi-directional solve \citep[e.g.][]{Kazemi2011,Smirnov2015} are viable solutions. In all the aforementioned cases, a good model for the surface brightness distribution of the A-team source is extremely valuable and, in many cases, essential. 

Recently, the detection of a broad absorption profile, centred at $78\pm1$~MHz in the sky-averaged signal has been reported by \cite{Bowman2018}. This boosted the interest in the ultra-low-frequency regime, driven by the possibility to detect neutral hydrogen during the cosmic dawn ($z\sim30$ to 15) and possibly even into the Dark Ages ($z\sim200$ to 30). The largest complication in these experiments is the subtraction of the strong astrophysical and instrumental foregrounds. The Galactic plane and the A-team sources are major contributors to the astrophysical foreground and a good model of these sources is paramount for their removal. Low-frequency, wide-field surveys have also renewed the interest of the broader scientific community \citep[e.g.][]{Shimwell2016a,Intema2017,Hurley-Walker2017}. For example, tracing cosmic rays (electrons) to the lowest energies provides insight into their inefficient acceleration mechanisms \citep[e.g.][]{deGasperin2017}. Low-frequency radio surveys can detect active galaxies in their late stages \citep[e.g.][]{bgmv16}, radio haloes and radio cluster shocks in merging clusters \citep[e.g.][]{Hoang2017}, and also the highest redshift radio sources \citep[e.g.][]{Saxena2018}. Again, our ability to carry out such surveys is limited by the extent that we can remove the contaminating emission from the bright A-team sources.

With the aim of determining accurate models for the surface brightness distribution of the A-team sources at low radio frequencies, we have carried out an imaging campaign with the Low Band Antenna (LBA) system of LOFAR, using the Dutch array. In Sec.~\ref{sec:observations}, we describe the observations of the four sources and in Sec.~\ref{sec:datareduction} we discuss the data reduction. In Sec.~\ref{sec:models}, we describe the models that we are releasing to the astronomical community, and in Sec.~\ref{sec:conclusions} we briefly describe the main scientific outcome of this work.

\begin{table*}
\centering
\begin{threeparttable}
\begin{tabular}{lcccccc}
Source name & \multicolumn{2}{c}{Coordinates} & \multicolumn{3}{c}{Flux density (Jy)} & Size\tnote{a}     \\
           & RA (J2000) & DEC (J2000)                     & @ 50 MHz & @ 150 MHz & @ 1.4 GHz   & (arcmin) \\
Cassiopeia A (3C\,461)                & \hms{23}{23}{27.94} & \dms{+58}{48}{42.4} & 27104 & 9856 & 1768 & 7.4 \bigstrut[t]\\
Cygnus A (3C\,405)                    & \hms{19}{59}{28.35} & \dms{+40}{44}{02.1} & 22146 & 10713 & 1579 & 2.3 \\
Taurus A (3C\,144, M\,1, Crab Nebula) & \hms{05}{34}{31.97} & \dms{+22}{00}{52.1} & 2008  & 1368  & 829  & 7.9 \\
Virgo A (3C\,274, M\,87)              & \hms{12}{30}{49.42} & \dms{+12}{23}{28.0} & 2635  & 1209  & 212  & 15.0\\
\end{tabular}
\begin{tablenotes}
    \item[a] Largest angular size as measured from LOFAR images at 50 MHz.
\end{tablenotes}
\end{threeparttable}
\caption{The A-team: coordinates, flux densities, and sizes}\label{tab:targets}
\end{table*}

\begin{table*}
\centering
\begin{threeparttable}
\begin{tabular}{lcccccc}
Source       & Obs. date & Obs. length & Number of SBs & Resolution\tnote{a} & Rms noise & Dynamic\\
             &           & (h)         &               & (arcsec)   & (\mjybeam) & range\\
Cassiopeia A & 26-Aug-2015 & 16 & 244 & \beam{10}{7} & 11 & 7700 \\
Cygnus A     & 04-May-2015 & 11 & 242 & \beam{9}{6} & 40 & 18000 \\
Taurus A     & 03-Mar-2016 & 9 & 244 & \beam{11}{8} & 6 & 35000 \\
Virgo A      & 12-Apr-2017 & 8 & 202 & \beam{15}{12} & 5 & 18000 \\
\end{tabular}
\begin{tablenotes}
    \item[a] At the mean frequency of 54 MHz.
\end{tablenotes}
\end{threeparttable}
\caption{Observations and image parameters}\label{tab:observations}
\end{table*}

\section{Observations}
\label{sec:observations}

The LOFAR \citep{VanHaarlem2013} radio interferometer is capable of observing at very low frequencies (10 to 250~MHz). Each LOFAR station is composed of two sets of antennas: the LBA, which operates between 10 and 90 MHz, and the HBA, which operates between 110 and 250 MHz. Currently, LOFAR is composed of 24 core stations (CS; maximum baseline: $\sim 4$~km), 14 remote stations (RS; maximum baseline: $\sim 120$~km), and 14 international stations (IS; maximum baseline: $\sim 2000$~km, not used for this work).

For this paper, we took four separate LOFAR LBA observations, one for each A-team source. For these observations, we restricted our frequency range between 30 MHz and 77~MHz. Below 30 MHz, RFI quickly dominates over the signal, while above 70 MHz the LBA bandpass quickly drops. The datasets were divided into 244 sub-bands (SB) of 195.3 kHz bandwidth each. The time resolution of all datasets was 1 s and the frequency resolution was 64 channels per SB ($\sim3$~kHz). After radio frequency interference (RFI) excision \citep{Offringa2010a}, the visibility datasets were averaged down to 10 s and 1 channel per SB. Some of the SBs were removed after inspection of the data if RFI was visible. We carried out the observations in LBA\_OUTER mode, which uses only the outer half dipoles of each 96-antenna LBA field. This reduces the field of view to a full width at half maximum (FWHM) of $\sim 4$ degrees at 60~MHz, and ignores the central dipoles where mutual coupling and un-modelled large-scale emission from the Galaxy make their calibration challenging. A summary of the observation parameters is given in Table~\ref{tab:observations}.

\section{Data reduction}
\label{sec:datareduction}

The data reduction follows roughly the strategy that has been outlined by \cite{deGasperin2019}, which was designed for point-like calibrator sources using the LBA system of LOFAR. All of our targets can also be considered bright calibrators, but the main difference is the complexity of their structure on $\sim10\arcsec$ to arcminute scales. To compensate for this, we had to rely on a large number of self-calibration cycles to reconstruct the morphology of the sources.

\subsection{Initial model and flux scale}

The initial model for the self-calibration was taken from the literature or from archival data. Each model was rescaled to match the expected integrated flux density for a given frequency. The integrated flux density is modelled following \cite{Perley2017},
\begin{equation}
\log(S[{\mathrm{Jy}}])=a_0+a_1\log(\nu[{\mathrm{GHz}}])+a_2[\log(\nu[{\mathrm{GHz}}])]^2+...,
\end{equation}
where $\nu$ is the frequency and $A_i$ a set of coefficients. At these low frequencies Faraday depolarisation is very efficient, therefore all models are unpolarised. We now explain how we build up the initial model for each target.

\begin{description} 
 \item[Cassiopeia A:] As a starting model, we used the LOFAR LBA image produced by \cite{Oonk2017}. The model was rescaled to match the \cite{Perley2017} flux density using the parameters they derived as follows: $a_0 = 3.3584$, $a_1 = -0.7518$, $a_2 = -0.0347$, and $a_3 = -0.0705$. We note that the flux density of Cassiopeia A decreases with time \citep[][and references therein]{Baars1977,Vinyaikin2014}.\\
 
\item[Cygnus A:] The initial model was taken from \cite{McKean2016} who observed this source using the LOFAR HBA system at frequencies between 109 MHz and 183~MHz that have an angular resolution of 3.5\arcsec. The model has a higher resolution than what is needed to start our self-calibration process, and the source is known to undergo a rapid turnover in the bright hotspots below 100 MHz \citep{McKean2016}. This makes the extrapolation of the HBA model just an approximation of the expected emission at LBA frequencies. The flux scale for Cygnus A has been estimated following \cite{Perley2017}. The best fit is a polynomial function of the fifth order with parameters $a_0 = 3.3498$, $a_1 = -1.0022$, $a_2 = -0.2246$, $a_3 = 0.0227$, $a_4 = 0.0425$.\\

\item[Taurus A:] There was no prior model available for this object. However, this source has a compact bright component (the pulsar at the centre of the supernova remnant) that provides $\sim10\%$ of the total flux density, or about 300 Jy at 50 MHz. We therefore started the self-calibration process assuming a point source model at the field centre and using only the shortest baselines (so that the entire source was seen as a point source) or the longest baselines (so that the extended component was resolved out and only the emission from the pulsar dominated the visibilities). In this way, we could obtain initial phase solutions for the LBA stations, which we then used to reconstruct the extended component of the source and continue the self-calibration process. The final model, with all of the components, was rescaled to match the \cite{Perley2017} scale using the parameters $a_0 = 2.9516$ and $a_1 = -0.2173$, $a_2 = -0.0473$, and $a_3 = -0.0674$.\\

 \item[Virgo A:] As a starting model, we used the low-resolution LOFAR LBA image presented by \cite{deGasperin2012}. The flux scale was set using a second order polynomial function with parameters $a_0 = 1226$, $a_1 = -0.8116$, and $a_2 = -0.0483$.
\end{description}

\begin{figure*}
\centering
\includegraphics[width=\textwidth]{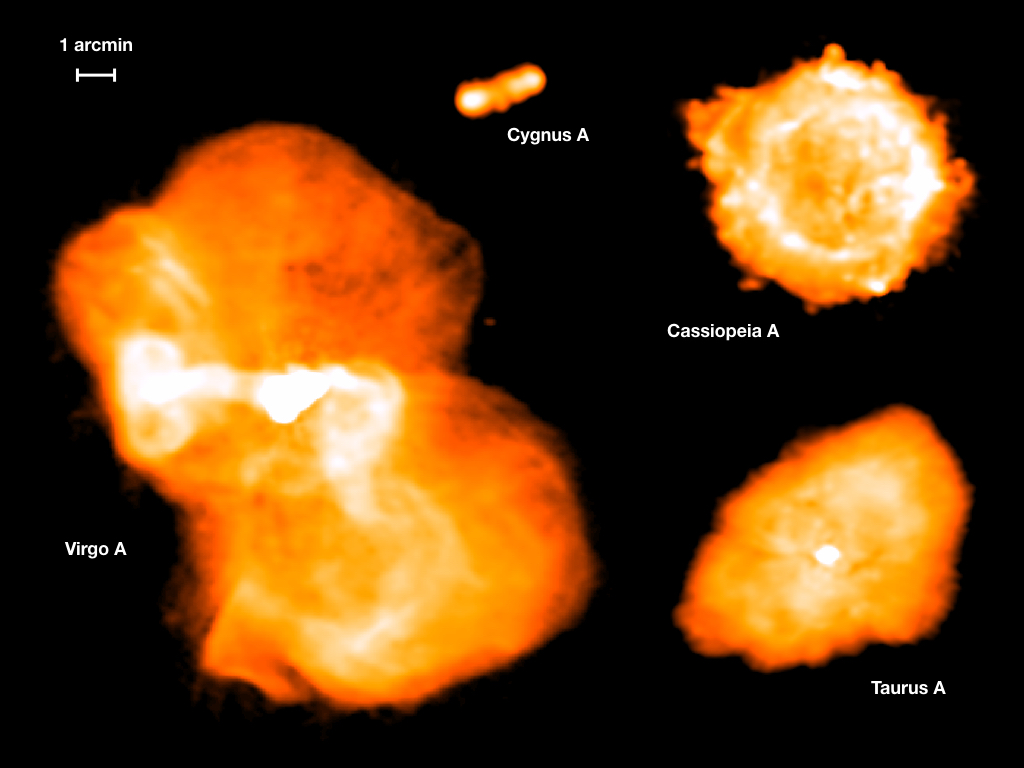}
\caption{Images of Cassiopeia A, Cygnus A, Taurus A, and Virgo A at a frequency of 50 MHz (using a bandwidth 30 to 77 MHz). Sources are scaled to show the correct apparent size ratio. The rms noise and resolution of each image are given in Table~\ref{tab:observations}.}\label{fig:ateam}
\end{figure*}

\subsection{Calibration}

The calibration procedure for all targets is described following the radio interferometer measurement equation (RIME) formalism \citep{Hamaker1996a,Smirnov2011a}. First, all of the data points on baselines shorter than $30 \lambda$ were flagged to remove any extended structure associated with the Galactic plane. We also retained only the part of the observations where the targets were above {15\deg} elevation. Then, a first round of (direction independent) calibration was performed. Initially, for each SB we solved for a diagonal and a rotational matrix simultaneously, so that the Faraday rotation effect is channelled into the rotational matrix, while all other effects remain in the diagonal matrix. The latter was then used to compare the XX and YY solutions (the two diagonal elements of the matrix) and to extract from the phases the differential delay between the two polarisations. This effect was then applied together with the element beam model of the LOFAR LBA \citep{VanHaarlem2013}. The data were then converted into a circular polarisation basis. In this basis, the effect of Faraday rotation can be described by a phase-only diagonal matrix with an opposite sign on the two circular polarisations. We solved per SB for a diagonal matrix and for each time step we fit the $\propto \nu^{-2}$ Faraday rotation effect on the difference between the two diagonal elements RR and LL. The dataset was then converted back to linear polarisation and corrected for Faraday rotation. Finally, a last diagonal matrix solve was performed at high frequency and time resolution to correct for ionospheric delay, clock errors, and the bandpass amplitude. These corrections were then applied and the dataset was ready for imaging and deconvolution.

\subsection{Imaging}

The imaging procedure for each self-calibration iteration was similar for all four targets. We used \texttt{WSclean} \citep{Offringa2014} to perform the deconvolution. We weighted the visibility data using a \citet{Briggs1995} weighting of $-1$ for Virgo A, $-1.2$  for Cassiopeia A and Taurus A, and $-1.4$ for Cygnus A. We chose these negative values to compensate for the large number of short baselines generated by the dense core of LOFAR. We used different weighting schemes to sample the different large and small scales of our targets. In all cases, we used multi-scale Clean with a large number of truncated Gaussian components, with scales up to the source extent. During imaging, the datasets were divided into 61 frequency blocks and imaged separately. All 61 images were combined to search for the peak emission to subtract during minor cycles. When the location of the clean component was determined, the brightness for that pixel was found for each image and a fourth order polynomial function was fitted through those measurements. These ``smooth'' components were then added to the model. The final images are shown in Fig.~\ref{fig:ateam}. The resolution of Cygnus A is higher than for the other sources to trace the more complex and compact structure for the source. However, the increase weighting of the data from the isolated LOFAR remote stations has an effect on the rms noise that, in this case, is four or more times higher than for the other sources. We did not perform any primary beam correction because, given the total extent of the sources, the average primary beam effect, even at the edges of our largest source, Virgo A, was always negligible ($<1\%$). In all cases, the target angular resolution for the models was achieved ($\theta_{\rm res}<15\arcsec$).

\section{Models}
\label{sec:models}

\begin{table*}
\centering
\footnotesize
\begin{threeparttable}
\begin{tabular}{lcccccccc}
Name & Type & Ra & Dec & $I$ & SpectralIndex & Ref.Frequency & MajorAxis & MinorAxis\\
 &  &  &  & (Jy) &  & (Hz) & (arcsec) & (arcsec)\\
s0c0 & POINT & 05:34:32.65 & 21.57.16.2 & 0.141 & [-0.018, 0.066, 1.504, -0.762] & 55369567 &--&--\\
s1c1 & GAUSSIAN & 05:34:23.88 & 22.03.22.1 & 1.473 & [-0.945, 1.228, 1.427, -12.222] & 55369567 & 70.644 & 70.644
\end{tabular}
\end{threeparttable}
\caption{Two example lines from the clean component list files. The ``Orientation'' column (not shown) is always set at 0\deg.}\label{tab:model}
\end{table*}

With this paper we provide the highest resolution models of the four A-team sources Cygnus A, Cassiopeia A, Taurus A, and Virgo A at ultra-low radio frequencies. The models are given in the on-line material in two different formats that are compatible with \texttt{WSclean} \citep{Offringa2014}. The first is a set of model FITS files including the clean components at 61 different frequencies, equally divided in the frequency range from 30 MHz to 77~MHz. The second is a text file including a list of clean components; the associated spectral shape is described by a seventh order polynomial function for Cygnus A, Cassiopeia A, and Taurus A, and by a fifth order polynomial function for Virgo A (see Table~\ref{tab:model}). Each clean component is one line of the file\footnote{The data format is explained in detail at \url{https://sourceforge.net/p/wsclean/wiki/ComponentList/}.}. Some aspects to note: the type of clean component can only be ``POINT'' (for point-like components) and ``GAUSSIAN'' for extended components. In the second case, the  MajorAxis and MinorAxis are saved to represent the FWHM of the component. The $I$ column represents the flux density in Jy at the reference frequency. The SpectralIndex column shows the coefficients of the  polynomial function when normalised to the reference frequency. The polynomial function is given by
\begin{equation}
 S_\nu = I + C_0\ (\nu/\nu_0 - 1) + C_1\ (\nu/\nu_0 - 1)^2 + ... ,
\end{equation}
where $I$ is the Stokes total intensity value, $\nu_0$ is the reference frequency, and $C_0$, $C_1$, ... are the coefficients saved in the SpectralIndex column. The $-1$ within round brackets is necessary to let the assumed Stokes $I$ be the correct value at the reference frequency. Currently, all Gaussian clean components are circular, that is, the MajorAxis and MinorAxis are the same. We also provide a low-resolution model in text-file format obtained by re-imaging the data at 45\arcsec resolution. These models have fewer clean components and can be efficiently used in arrays with a more compact configurations.

With these models the A-team sources can also be used as calibrators for ultra-low-frequency observations. However, if the sources are strongly resolved, then the flux density on the longest baselines might not be enough. In these cases, fainter but more compact sources such as 3c\,196, 3c\,380, or 2c\,295 are preferred.

\section{Discussion and conclusions}
\label{sec:conclusions}

We obtained data for the four radio sources with highest flux density in the northern sky using the LOFAR LBA system (Dutch array). We release these high-fidelity, high-resolution models of these sources in the frequency range 30 to 77~MHz. A detailed analysis of each source is beyond the scope of this paper, and will be carried out in separate individual publications for each object. Nonetheless, in this section, we report an overview of our findings.

\begin{description}
 \item[Cassiopeia A:] LOFAR LBA data for Cassiopeia A, a $\sim 330$~yr old supernova remnant, has been analysed recently by \citet{Arias2018}. However, we note that the data and reduction methods presented in this work are new. The most striking feature of Cassiopeia A at low radio frequencies is the effect of internal free-free absorption from cold \citep[$\sim 100$~K;][]{Arias2018,Oonk2017}, unshocked supernova ejecta material. As a result, the central region of Cassiopeia A (interior to the bright shell) is less bright than in the gigahertz band. The presence of such internal absorption was first noted by \citet{Kassim1995}, and further investigated by \citet{DeLaney2014} with images down to 74 MHz.

The overall flux density within a beam centred on Cassiopeia~A is always affected by free-free absorption by relatively cool free electrons between us and the source, as well as internal absorption in the central region. As a consequence, for a given beam, the flux density can be described as \citep{Arias2018}
\begin{equation}
S_\nu = S_0\left(\frac{\nu}{\nu_0}\right)^{-\alpha} \left[f + (1-f)\,\textrm{e}^{-\tau_{\nu,\rm int}}\right] \textrm{e}^{-\tau_{\nu, \rm ISM}},
\end{equation}
where $f$, the flux fraction, comes from the unobscured part of the shell, and $(1-f)$ the covering fraction (i.e. the back side of the supernova-remnant shell); $\tau_{\nu,\rm int}$ is the optical depth due to free-free absorption from the unshocked ejecta, and $\tau_{\nu,\rm ISM}$ is the free-free absorption due to the free electrons between us and Cassiopeia A. The free-free absorption scales as $\tau_\nu \propto \nu^{-2}T^{-3/2}n_\mathrm{e}\sum_i n_i$, which shows that the internal mass estimate is dependent on the temperature of the free electrons and the composition and degree of ionisation of the unshocked supernova ejecta. Moreover, clumping of the ejecta may seriously affect the relation between the internal, unshocked mass, and the internal free-free absorption.

The effect of the internal absorption is that the central part of Cassiopeia A is less bright below 100~MHz than at high frequencies. Once $\tau_{\nu,\rm int} \gg 1$ the total flux density continues to fall again as $S_{\nu}\propto \nu^{\alpha}$, but with a flux reduced by $(1-f)$ compared to the extrapolation from high frequencies, except that the external free-free absorption causes an overall reduction of the flux density. As a result, the maximum flux density of Cassiopeia A occurs around 20~MHz \citep{Baars1977}.

The multi-channel LOFAR LBA data provide a more precise localisation of the effect and infer an unshocked ejecta mass of $(3\pm0.5)~{\rm M}_\odot$. We plan to update this result using the new calibration and data reduction procedures presented in this work (Arias et al. in prep.).
 
 \item[Cygnus A:] This is the first work examining this source both at this frequency and resolution. We report the work of \cite{Lazio2006}, which reached a similar resolution of our LOFAR images at 74 MHz. The most striking feature in the new LOFAR LBA image of Cygnus A is the absence of hotspots that are seen at higher frequencies \cite{McKean2016}. After convolving all of the LBA images to the same resolution, we attempted the extraction of the in-band spectral index $\alpha$ (with $s_\nu \propto \nu^\alpha$) in three regions with a size that is equivalent to the convolved beam. We positioned two regions close to the east and west edges of the source, and they gave spectral index values of $\alpha=0.46\pm0.05$ and $\alpha=0.25\pm0.05$, respectively, for hotspots A and D \cite[defined in][see Fig.~\ref{fig:cyg}]{McKean2016} between 30 MHz and 77 MHz. The third region was positioned at the source centre, close to the southern plume, which gave a spectral index of $\alpha=-0.82\pm0.05$ between 30 and 77 MHz. We note that beam dilution almost certainly biases the results towards steeper values in the case of the hotspots. To calculate the uncertainties, a conservative flux error of 10\% in each measurement was added to the error estimated from the map noise.\\
 
\begin{figure}
\centering
\includegraphics[width=.5\textwidth]{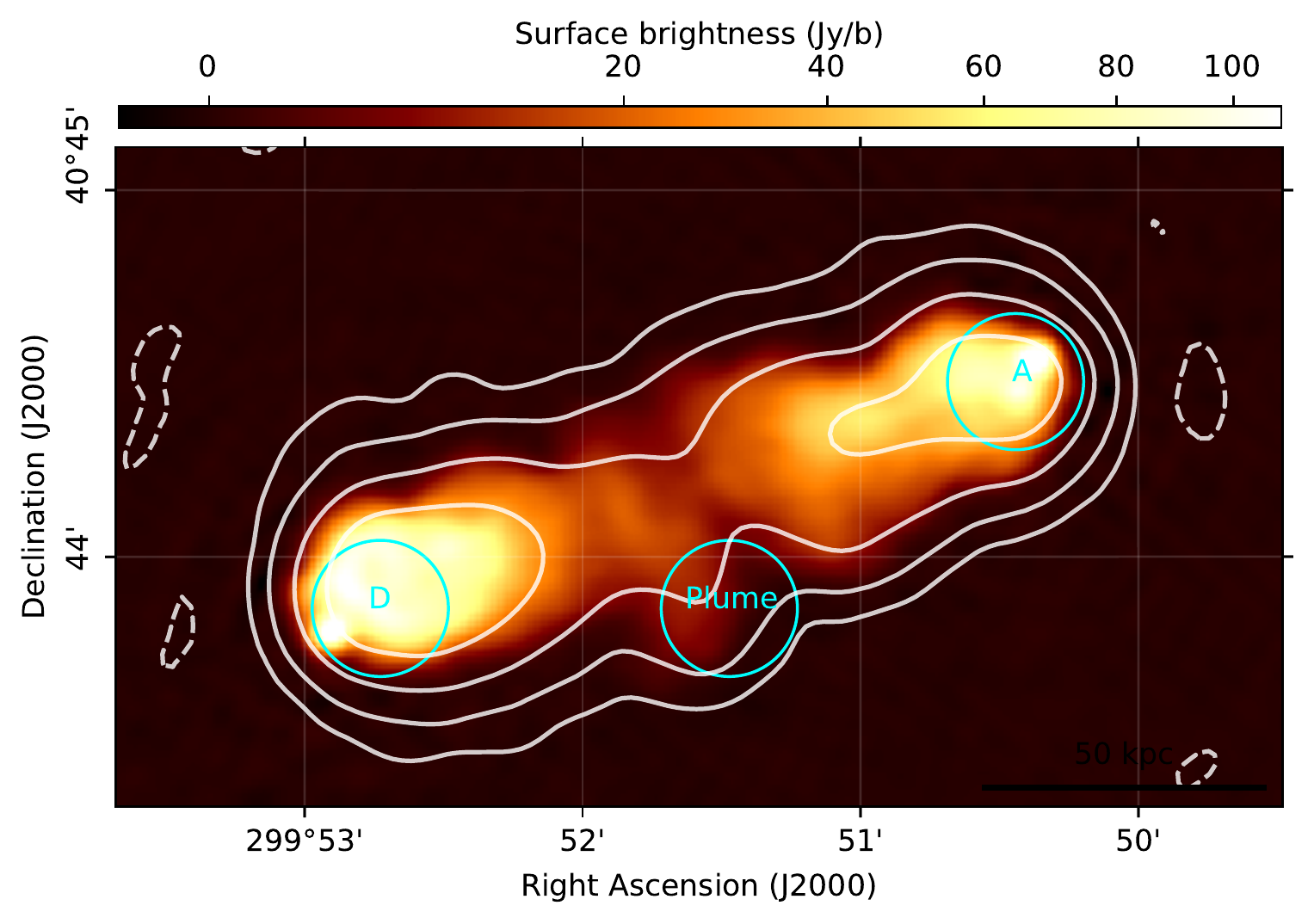}
\caption{LOFAR HBA image of Cygnus A (central frequency of 146 MHz). The resulting rms map noise is 43~\mjybeam and the FWHM beam size is \beam{3.8}{2.7} \citep[from][]{McKean2016}. Contours from the LBA map at $-0.3$, 10, 30, 100, 300 Jy beam$^{-1}$ are superimposed. The circles represent the regions where we extracted the LBA in-band spectral index.}\label{fig:cyg}
\end{figure}
 
As discussed by \cite{McKean2016}, the spectral energy distribution in the hotspot regions A and D peaks between 140 and 160~MHz, and then starts decreasing towards lower frequencies. The two main models proposed to explain the turnover are as follows: (i) free-free absorption or synchrotron self-absorption processes within the hotspots or along the line of sight \citep{Kassim1989} and/or (ii) a cut-off in the electron energy distribution at low energies \citep{Carilli1991}. From the LOFAR HBA data, in combination with higher frequency data from the VLA, \citet{McKean2016} found that the strong turnover in the spectral index ruled out a cut-off in the electron energy distribution at low energies, and the limit in the spectral index provided by the new LOFAR LBA imaging is consistent with that conclusion. \citet{McKean2016} also found that the synchrotron self-absorption model was also unlikely, given the very large magnetic field strengths needed to cause such a turnover ($B \approx 1$ to 2~G) relative to the modest magnetic field strength that is required by the synchrotron cooling model for both hotspots ($B \approx 150~\mu$G). The free-free model was also challenging to explain the data since the implied electron densities ($n_e \approx 2$~cm$^{-3}$) should result in a significant de-polarisation of the emission seen at GHz frequencies, which is not the case. Only low-frequency observations of the hotspots in the LOFAR LBA can distinguish between different models. However, our resolution is not sufficient to constrain such models, and therefore observations with the international baselines, to achieve the arcsecond resolution needed, are planned. Nevertheless, our inverted spectrum for the hotspot regions confirms that some form of absorption must be at least partially responsible for the observed turnover.\\
 
The plume extending from the central part of the source towards the south is also visible and the in-band spectral index is in line with what is measured at higher frequencies by \cite{McKean2016}. Finally, we report the detection of diffuse emission, with an extension $\sim 4\arcmin$ towards the north-east of Cygnus A. The classification of this source is difficult because of the dynamic range of the image. It could be a background radio galaxy or some emission related to the intra-cluster medium dynamics.\\
 
\item[Taurus A:] This radio source is associated with the Crab Nebula \citep[see][for a review]{Hester2008,Buehler2013}, which is the supernova remnant of SN\,1054 \citep[e.g.][]{Stephenson2002}. However, most of the electromagnetic radiation is coming from the pulsar wind nebula (PWN) that is powered by the Crab Pulsar (PSR B0531+21), which has a period of 33~ms, and a rotational energy loss rate of $\dot{E}=4.6\times 10^{38}$~erg\ s$^{-1}$. Taurus A is unique in that synchrotron emission is dominating the spectrum from low radio frequencies up to $\sim 100$~MeV ($\sim 10^{22}$~Hz). Synchrotron emission even dominates the optical and UV band \citep{Veron-cetty1993}, but the optical also reveals strong line emission from the filaments of ionised supernova ejecta. The radio synchrotron spectrum has a spectral index of $\alpha \approx -0.3$ \citep{Green2019}, but in the optical the spectral index is closer to $\alpha \approx -0.8$ \citep{Veron-cetty1993} and it is even steeper in X-rays with $\alpha \approx -1.1$ \citep{Madsen2017}. The spectral break between the radio and optical can be understood as due to synchrotron cooling, giving a magnetic field of $B\approx 100$ to 200~$\mu$G and an age of $\sim 950$~yr, but the steep X-ray spectrum is not well understood. One suggestion is that there are two populations of relativistic electron/positron: one responsible for the radio emission and another for the UV/X-ray emission \citep[e.g.][]{Meyer2010}. The radio population could be the result of a past injection of particles ;\citep[i.e. ``relic electrons/positrons''][]{Atoyan1996}, or two different electron/positron acceleration mechanisms, such as reconnection for the low-energy population associated with the radio emission, and diffusive shock acceleration for the higher energy particles responsible for X-rays. To complicate things, the injection of fresh electron/positrons seems to occur on the inside of the bright optical/X-ray torus \citep[][and reference therein]{Hester2008}, but some X-ray emission is also associated with two jets that are roughly orientated south-east to north-west.\\

The radio emission from Taurus A in the LOFAR LBA, as seen in Fig.~\ref{fig:ateam}, is elongated, in the south-east to north-west direction. This is similar to higher frequency maps \citep[e.g.][for a 5.5~GHz VLA map]{Bietenholz2015}. However, what is at least qualitatively different between the low- and high-frequency radio maps is that at low frequencies there seems to be relatively less emission from the torus region and more emission associated with the "jets", suggesting that these two components have different spectral indices, which could potentially shed new light into whether the PWN consists of a single electron/positron population with a complicated energy distribution or  two or even more populations with different physical origins. We caution, however, that this needs to be further investigated as the dynamical range and the uv-coverage of the LOFAR LBA and 5.5 GHz VLA maps are not similar, requiring care to assess quantitative differences. We will come back to this issue in a future paper dedicated to the LOFAR LBA observation of Taurus A presented in this work.\\

Finally, we note that the centre of Taurus A is dominated by the emission from the steep spectrum of pulsar with an in-band spectral index of $\alpha= -1.50 \pm 0.05$, in line with previous measurements \citep{Bridle1970}.\\

\item[Virgo A:] This is the most extended of the A-team sources, reaching an apparent scale of about 15\arcmin. Virgo A is the radio emission associated with the active galaxy M\,87 and is famous for hosting one of the best-studied supermassive black holes \citep[recently imaged by][]{Akiyama2019}. The central cocoon, which at these frequencies accounts for just $\sim30\%$ of the total source flux density, hosts the well-known one-sided jet and morphologically resembles an FR\,II radio galaxy. However, Virgo A emission extends well beyond the central cocoon and the majority of the flux density comes from a relatively low-surface brightness envelope filled with filamentary structures. In this region, clear connection between the radio and the X-ray emission shows one of the best examples of active galactic nucleii feedback in action, where cold gas is uplifted by buoyantly rising bubbles towards the outskirts of the galaxy potential well \citep{Forman2007}. The external boundaries of the source appear well confined even at ultra-low frequencies; this was already observed at higher frequencies \citep{Owen2000,deGasperin2012}. The resolution of these new maps will enable the first detailed spectral study of the source envelope and of the embedded filamentary structures. This analysis will be part of a future publication.
 
\end{description}

\begin{acknowledgements}

The Leiden LOFAR team gratefully acknowledge support from the European Research Council under the European Unions Seventh Framework Programme (FP/2007-2013)/ERC Advanced Grant NEWCLUSTERS-321271.

AB acknowledges financial support from the Italian Minister for Research and Education (MIUR), project FARE SMS, code R16RMPN87T and from the ERC-Stg DRANOEL, no 714245.

LOFAR, the Low Frequency Array designed and constructed by ASTRON, has facilities in several countries, that are owned by various parties (each with their own funding sources), and that are collectively operated by the International LOFAR Telescope (ILT) foundation under a joint scientific policy.
This research has made use of NASA's Astrophysics Data System.

\end{acknowledgements}


\bibliographystyle{aa}
\bibliography{library}


\end{document}